\documentclass[12pt]{JHEP}
\usepackage{epsfig}

\def\fun#1#2{\lower3.6pt\vbox{\baselineskip0pt\lineskip.9pt
  \ialign{$\mathsurround=0pt#1\hfil##\hfil$\crcr#2\crcr\sim\crcr}}}
\newcommand{\beq}{\begin{equation}}
\newcommand{\eeq}{\end{equation}}

\title{Signatures of galactic magnetic lensing\\
upon ultra high energy cosmic rays}

\author{Diego Harari$^a$, Silvia Mollerach$^b$ and Esteban Roulet$^b$
\\$^a$Departamento de F\'\i sica, FCEyN, Universidad de Buenos Aires
\\Ciudad Universitaria - Pab. 1, 1428, Buenos Aires, Argentina
\\$^b$Departamento de F\'\i sica, 
Universidad Nacional de La Plata\\ CC67,  1900, La Plata, Argentina
\\ Email: \email{harari@df.uba.ar, mollerac@venus.fisica.unlp.edu.ar, 
roulet@venus.fisica.unlp.edu.ar}}

\abstract{We analyse several implications of lensing by the regular 
component of the galactic magnetic field upon the observed properties of 
ultra high energy cosmic rays. Magnetic fields deflect cosmic ray 
trajectories, causing flux (de)magnification, formation of 
multiple images of a single source, and time delays. We derive 
the energy dependence of these effects near the caustics at which 
the flux amplification of a point source diverges.
We show that the large magnification of images
around caustics leads to an amplification bias, which can make
them dominate the flux in some energy ranges.
We argue that clustering in the arrival directions
of UHECRs of comparable energy may be due to magnetic lensing 
around caustics. We show that magnetic lensing 
can also significantly alter the observed composition of cosmic rays at 
the highest energies. We also show that the time delay between events 
from a single image may monotonically decrease with decreasing 
energy in the neighborhood of a caustic, opposite to its behaviour in 
normal regions.}

\preprint{.}

\begin{document}

\section{Introduction}

Galactic and intergalactic magnetic fields deflect extragalactic
charged cosmic ray trajectories in their journey from their sources 
to the Earth, causing several effects upon the observed properties of
ultra high energy cosmic rays (UHECRs). For a review see for instance 
\cite{sigl98}.

In a previous paper \cite{I} (in what follows paper I), we have shown 
that the regular component of the galactic magnetic field 
acts as a giant lens upon charged CRs, and this can sizeably 
amplify (or demagnify) the flux arriving from any given source,
modifying its spectrum.
The magnification of the CR flux by the galactic magnetic field 
becomes divergent for directions along critical curves in the sky 
seen from the Earth, corresponding to caustic curves in the 
``source plane'', i.e. in the corresponding directions outside 
the Galaxy. The location of the caustics move with energy and, 
as a caustic crosses a given source direction, pairs of additional 
images of the source appear or disappear. Multiple image formation
is a rather common phenomenon, at least within the 
galactic magnetic field models considered in I. Indeed, the caustics
sweep a rather significant fraction of the sky as the ratio $E/Z$
between the CR energy and charge
steps down to values of the order of a few EeV ($1~{\rm EeV}= 
10^{18}~{\rm eV}$), before the drift and 
diffusive regimes turn on. At values of $E/Z$ larger than around 50 EeV  
caustics are present but sweep out a relatively small fraction of the 
sky.  The effects under discussion are thus relevant even for the
highest energy events so far detected if the CRs have a component
which is not light.

In the present paper we analyse in detail the energy dependence of the
magnification of image pairs near a caustic, and discuss several
implications of the existence of caustics upon the observed properties 
of UHECRs. 

In section \ref{caustics} and in  appendix A we show that if a source 
lies along a caustic at energy $E_0$, the magnification 
factor $\mu$ of each image in the pair that becomes visible at
energies below $E_0$ diverges as $\mu\approx A/\sqrt{1-E/E_0}$. $A$ is a 
dimensionless constant, fixed by the structure of the magnetic 
field along the CR trajectory to the source.  It must be determined
numerically, and we do so for some examples of source locations.
We discuss several implications of this result, such as
the expected enhancement in the observed event rate at energies near
a caustic, which can lead to an amplification bias, making it more likely
to detect images of UHECR sources that lie along caustics than
sources in ordinary regions. 
In appendix B we provide a simple proof of the relation between
the magnification of the three images appearing when there are
two nearby folds.  In section \ref{angles} we show that the angular 
separation between images in a pair increases near the caustic 
as $\Delta\theta\propto\sqrt{1-E/E_0}$. We analyse the expected event 
rate from the original source and its multiple images not only as 
a function of energy but also in terms of the observed arrival 
direction. We show that magnetic lensing 
near caustics is a source of clustering 
in the arrival directions of events with comparable energy.
In section \ref{composition} we stress the fact that at fixed 
energy the magnetic lensing effects depend upon the CR electric charge, 
and show that this dependence can significantly alter the observed CR 
composition at the highest energies.
In section \ref{td} we discuss features of the time delay between 
events from multiple images, relevant in the case of bursting or 
highly variable sources of UHECRs \cite{wax96}. We show that 
the delay between the arrival of events of equal energy from the 
two images in a pair increases as $\Delta t\propto\sqrt{1-E/E_0}$ 
around a caustic. We also show that the
time delay between events at different energy from one image 
in a pair 
may monotonically decrease with decreasing energy near the
caustic, opposite to the behaviour of time delays in normal 
regions.
Section \ref{conclusions} rounds up our conclusions.

The galactic magnetic field model that will be used throughout
this paper to illustrate magnetic lensing effects is the
bisymmetric spiral model with even symmetry (BSS-S) described in
paper I, which is a smoothed version of one of the models used in
\cite{st97} and \cite{me98} to study the effects of 
CR deflections in our galaxy. We refer the reader to paper I
for details about the field configuration and about the 
numerical methods implemented to determine CR trajectories
and flux magnifications.

\section{Flux enhancement by magnetic lensing near critical points}
\label{caustics}

As shown in paper I, the galactic magnetic field can act as
a giant lens that amplifies or demagnifies extragalactic sources
of UHECRs, much alike 
the gravitational lensing effect upon distant
quasars by intervening matter along the line of sight \cite{sch92}.
CRs from a distant extragalactic  source that
enter the galactic halo from a direction $(\ell,b)_{H}$ 
(in galactic coordinates) are deflected by the magnetic
field and thus observed on Earth as if coming from a different
direction $(\ell,b)_E$, and their flux is amplified (or demagnified)
by a factor $\mu$.
The magnitude of the effect depends upon the direction of observation
and upon the ratio $E/Z$ between energy and charge of the CR. 
The effect is most dramatic at the critical curves of the lens mapping,
the directions along which the observed flux diverges for a fixed
value of $E/Z$. The corresponding lines in the source coordinates
(the direction from which CRs enter the galactic halo)
are the caustics of the lens mapping. The location of the caustics
changes with energy, for fixed $Z$.

If a source position lies along a caustic of the 
magnetic lens mapping at energy $E_0$, a pair 
of images of the source either becomes visible or dissapears at 
energies below $E_0$. Their magnification diverges 
at $E=E_0$.
This behaviour was illustrated in paper I for sources at 
galactic coordinates
$(\ell,b)=(282.5^\circ,74.4^\circ)$ (M87 in the Virgo cluster)
and $(\ell,b)=(320^\circ,-30^\circ)$ (visible from the southern
hemisphere).  

In appendix A we give a geometrical interpretation of the magnification
of a pair of images near a caustic. Similar to the gravitational lens
case \cite{ch79,bl86,sch92b}, the magnification diverges as $1/\sqrt x$, 
where $x$ measures the distance of the source to the caustic. In the magnetic 
lensing case the location of the caustics ``move'' with energy,
and thus the magnification of a pair of images diverges as the 
energy approaches the energy at which the source lies along the
caustic as $1/\sqrt{E_0-E}$.  In appendix A we also find the first two
corrections
to this leading order behaviour, and show that the magnification of
the two images behaves as:
\begin{equation}
\mu_i(E)\approx{A\over \sqrt{1-E/E_0}}\pm B+C_i\sqrt{1-E/E_0}~.
\label{mu(E)}
\end{equation}
The dimensionless coefficient $A$ is the same for the two images.
The constant term $\pm B$ has the same value with opposite sign for
each image. The third term has a different coefficient $C_i$ for
each image due to the truncation of the expansion up to this order. 
The value of the coefficients $A, B$ and $C_i$ are fixed
by the properties of the magnetic field along the CR trajectories. 
We determine them through a fit
to the numerical output for the energy dependence of the amplification,
evaluated through the method described in paper I.

Figure~\ref{mufig} displays the numerical result for the amplification
near the caustics along with the fit to the analytic expression 
in eq. (\ref{mu(E)}), for the same examples of source locations as
in paper I. The top panel corresponds to M87 and the bottom panel
to the source in the southern hemisphere. The figures display the 
magnification of the principal image (the one that is also visible at 
the highest energies) and of the images $A$ and $B$, visible
at energies below the caustic only.
The analytic expression (\ref{mu(E)}) fits with very high accuracy the 
numerical result for the magnification of the secondary images near 
the caustic, at least down to energies 10\% below the energy $E_0$
of the caustic. The fit to M87 determines $E_0/Z= 20.41$~EeV,
$A=1.3$, $B=4.0$, $C_A=-3.1$, $C_B=5.0$. The fit to the southern 
source fixes  $E_0/Z =15.425$~EeV,
$A=0.44$, $B=0.37$, $C_A=-0.19$, $C_B=-0.28$.

\FIGURE{\epsfig{file=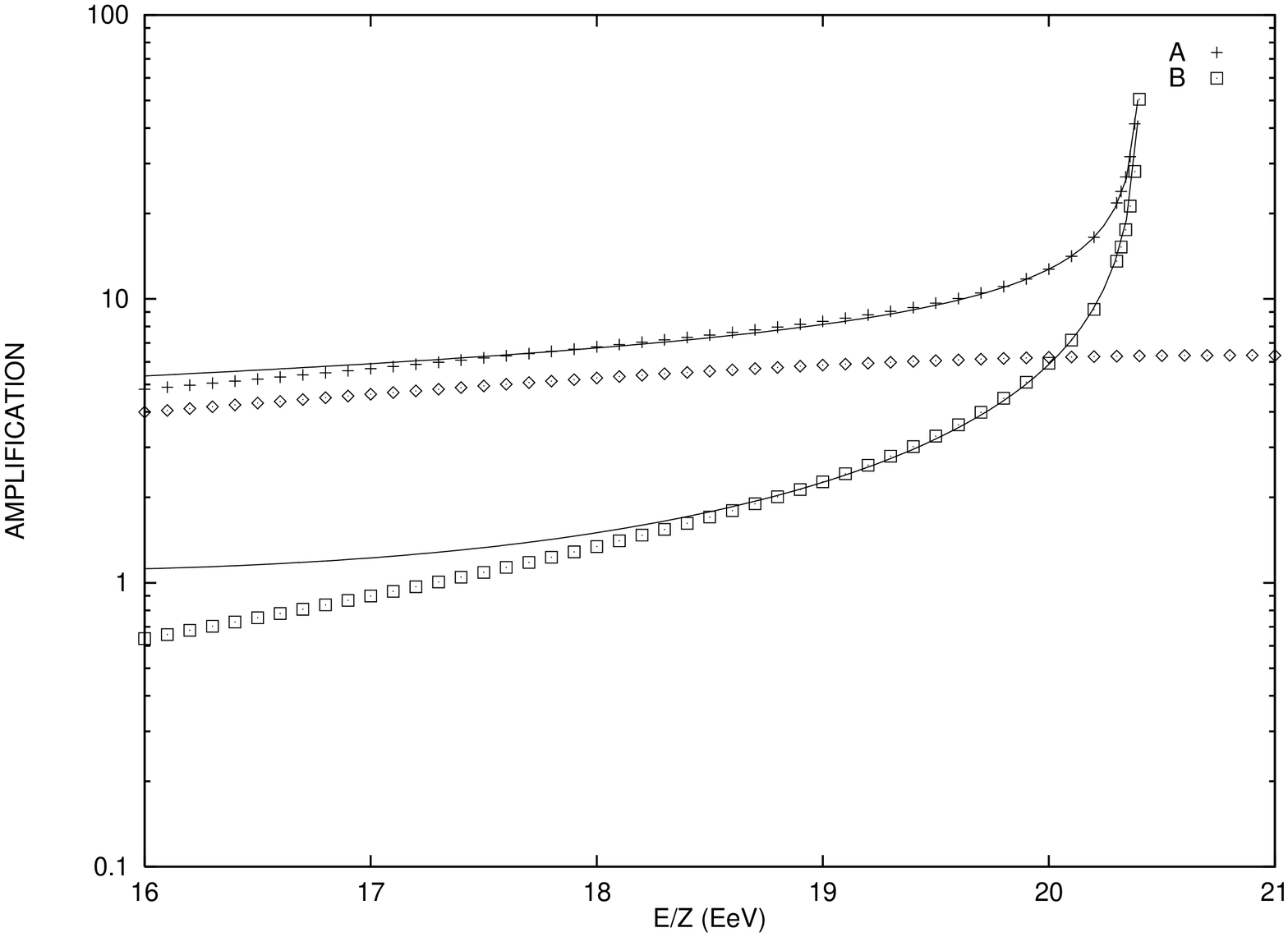,width=6cm,angle=-90}
\epsfig{file=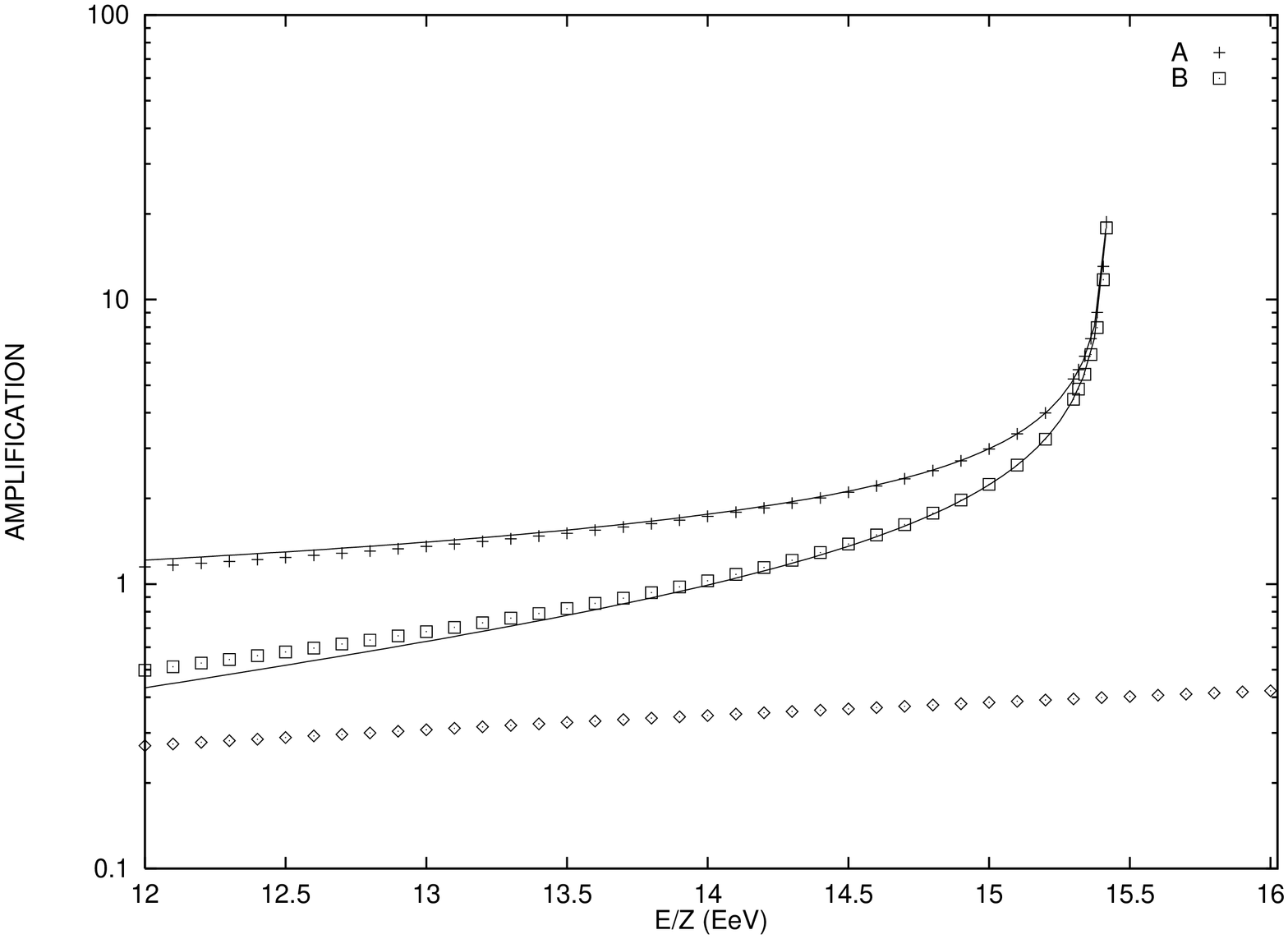,width=6cm,angle=-90}
\caption{Numerical result (points) and analytic fit to eq.~(\ref{mu(E)})
(solid lines) for the amplification near a caustic of a pair of images
of a point source. The source in the top panel is 
at $(\ell,b)=(282.5^\circ,74.4^\circ)$ (M87) 
and the source in the bottom panel is at $(\ell,b)=
(320^\circ,-30^\circ)$. The numerical result for the
amplification of each principal image (diamonds) is also shown.}
}\label{mufig}

The divergence of the flux magnification of a CR source at 
$E=E_0$ is softened down
to finite values when integrated across an extended source. 
This still allows for extremely large magnifications.
The limiting factor to the maximum attainable magnification 
in a realistic situation is not the extended nature of the sources,
but the fact that the divergence in the magnification, even in the 
case of a point source, arises only at a fixed energy $E_0$.
Since realistic sources are not monoenergetic, the integrated 
flux of a magnified source around $E_0$ is thus always finite, 
even in the point source approximation.

Large magnification of image pairs around caustics
leads to a significant enhancement of the detection
probability of a source in a flux-limited sample.  In gravitational
lensing this effect is termed ``amplification bias''\cite{tur84}, 
and may be responsible for the observation of quasars that  
would otherwise be too dim to detect. Let us now consider the
strength of this effect due to galactic magnetic lensing of UHECRs.
Consider a differential flux of CRs 
injected in the galactic halo at energies beyond the
``ankle'' given by $dF=F_0(E_0/E)^{2.7}dE$.
The flux observed on Earth coming from two images
formed at a caustic at energy $E_0$, in the energy
interval between $0.9~E_0$ and $E_0$, is 
\begin{equation}
F_{A+B}(0.9E_0<E<E_0)\approx 2\int_{0.9E_0}^{E_0}{\rm d}E\frac{{\rm
d}F}{{\rm d}E}
\frac{A}{\sqrt{1-E/E_0}}\approx 12 A \int_{0.9E_0}^{E_0}
{\rm d}E\frac{{\rm d}F}{{\rm d}E}~.
\label{enhancement}
\end{equation}
We have neglected a small $O(\sqrt{1-E/E_0})$ correction
(the term proportional to $C_A+C_B$).
The flux observed from the two images in this energy interval
near the caustic
is $12A$ times larger than the flux of the principal 
image of the source in the same energy range in the 
absence of magnification. This is also $2.4A$ times the 
flux that would arrive from the principal image at 
all energies above $E_0$
if there were no magnetic lensing. Detection of an UHECR source
in a narrow energy range around a caustic may thus be
more likely than its detection at any higher energy.
It may also be more likely to detect the CR source
at energies around the caustic than at significantly
lower energies (say half the energy of the caustic or
even less) in cases where both the principal image
as well as the secondary images are not magnified
at energies below the caustic. 

\FIGURE{\epsfig{file=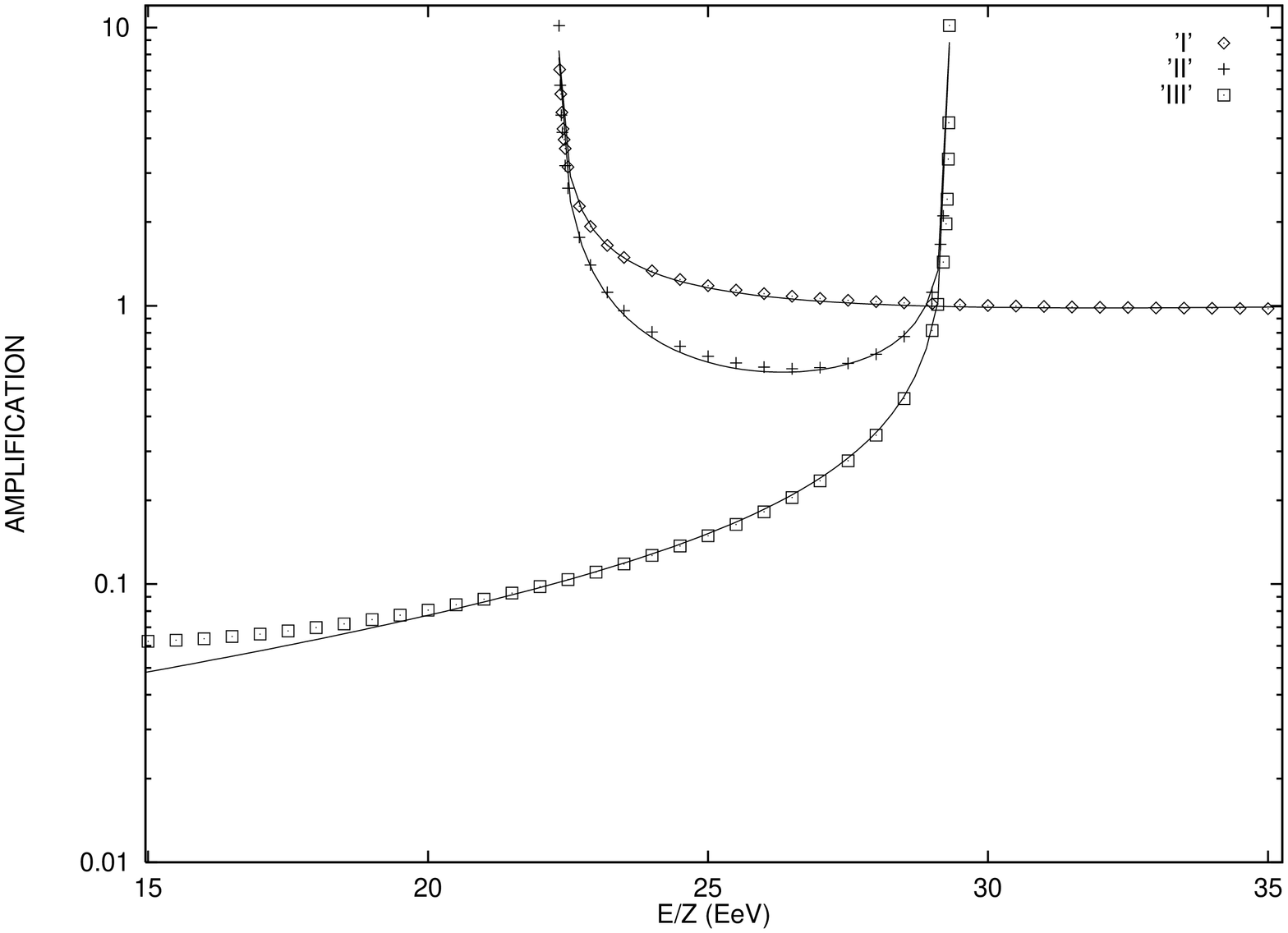,width=8cm,angle=-90}
\caption{Example of a source position that  is crossed by 
two folds at nearby energies. 
The solid lines for images I and III are the analytic fits of 
the numerical
results to eq.~(\ref{mu(E)}) for the amplification near a caustic.
The solid line for image II is $\mu_I +\mu_{III} - 0.7$.
The source is located at $(\ell,b)=(90^\circ,-10^\circ)$. 
}}\label{2foldsfig}

Large enhancements of the observed flux in a narrow
energy range also occur in the interesting
case when the source position lies along two caustics
at nearby energies so that the principal image of the source
(the only one visible from Earth at the highest energies) 
has also divergent magnification. Such a situation is
exemplified in figure~\ref{2foldsfig}, for a source located
at $(\ell=90^\circ,b=-10^\circ)$. A pair of images becomes 
visible at
energies below $E_0^{III}=29.31$~EeV, one of which 
disappears in a caustic along with the principal image 
at an energy $E_0^{I}=22.31$~EeV. The secondary image 
that merges with the principal image is that with opposite
parity.

In appendix B we develop the geometrical
interpretation and analytic approximation to the magnification
of images for two nearby folds. The analog result in the
gravitational lens case was obtained in \cite{sch92b}. 
As shown in appendix B, and 
verified numerically with high accuracy,
between the folds 
the sum of the magnifications of the two images with equal parity 
coincides with the magnification of the
image with opposite parity, up to a constant term.
The analytic fits in figure~\ref{2foldsfig} to the divergent
magnification of the principal image $(I)$ at $E_0^{I}$ and to the
magnification at $E_0^{III}$ of the image that survives at
low energies $(III)$, lead to $A_{I}=0.275$, $B_I=0.15$,
$C_I=0.63$;
$A_{III}=.096$, $B_{III}=-0.11$, $C_{III}=0.03$. 
The magnification of the image that is visible 
at energies between $E_0^{I}$ and $E_0^{II}$
is fit to $\mu_{II}=\mu_I+\mu_{III}-0.7$.

\section{Angular distribution of events: clustering around caustics}
\label{angles}

The arrival directions of the so far detected CRs in the
highest energy range is compatible (within the limited statistics
available) with an isotropic distribution, except for some small
angle clustering of events (eight doublets and two triplets within 
a total of 92 events with energies above 40 EeV)\cite{uchi99}. 
The observed relatively uniform distribution 
does not preclude the possibility that UHECRs originate from very
few nearby sources, if their trajectories underwent significant 
energy-dependent deflections in their journey to the Earth.
Trajectories of UHECRs with $E/Z$ below 
approximately 50 EeV  are sensibly deflected  in 
the magnetic field model considered in this paper. 
Several quite separated events may actually originate from the same 
source if CRs have a component which is not light (see paper I and 
references therein).
It has even been speculated \cite{ahn99} that all the events so
far detected at energies above $10^{20}$ eV may come from M87 in the
Virgo cluster, if the Galaxy has a rather strong and extended
magnetic wind.

In paper I we illustrated the angular displacement in the arrival
directions of CRs from the principal and secondary images
of a magnetically lensed source. Here we analyse in more detail
the angular displacement of image pairs around caustics. We argue that 
caustics can produce a significant  clustering of arrival directions 
of events with comparable energies.

It may well be the case (depending on the sources flux and 
composition, and on the exact nature of the galactic magnetic field) 
that a large fraction of the trajectories of the 
UHECRs so far detected have been significantly bent and do not point
to their sources.  In this scenario clustering of events would be
infrequent, at relatively low energies due to large deflections
and at high energies due to the smaller flux. However,
when a source is near a caustic, since the significant enhancement
of its flux occurs within a narrow energy range, the 
CRs arrive  from relatively nearby directions,
leading to an angular concentration of events.

We illustrate the clustering effect in figure~\ref{evfig}.
We consider the same examples of source locations as in
the previous section. We assume that the differential
flux injected by the source scales as $E^{-2.7}$, and that
the detecting system has the same efficiency at all energies within
the range considered. The energy range was divided in 50 bins of 
equal detection probability. Figure~\ref{evfig} displays
the predicted arrival directions of the events detected
from each source.

\FIGURE{\epsfig{file=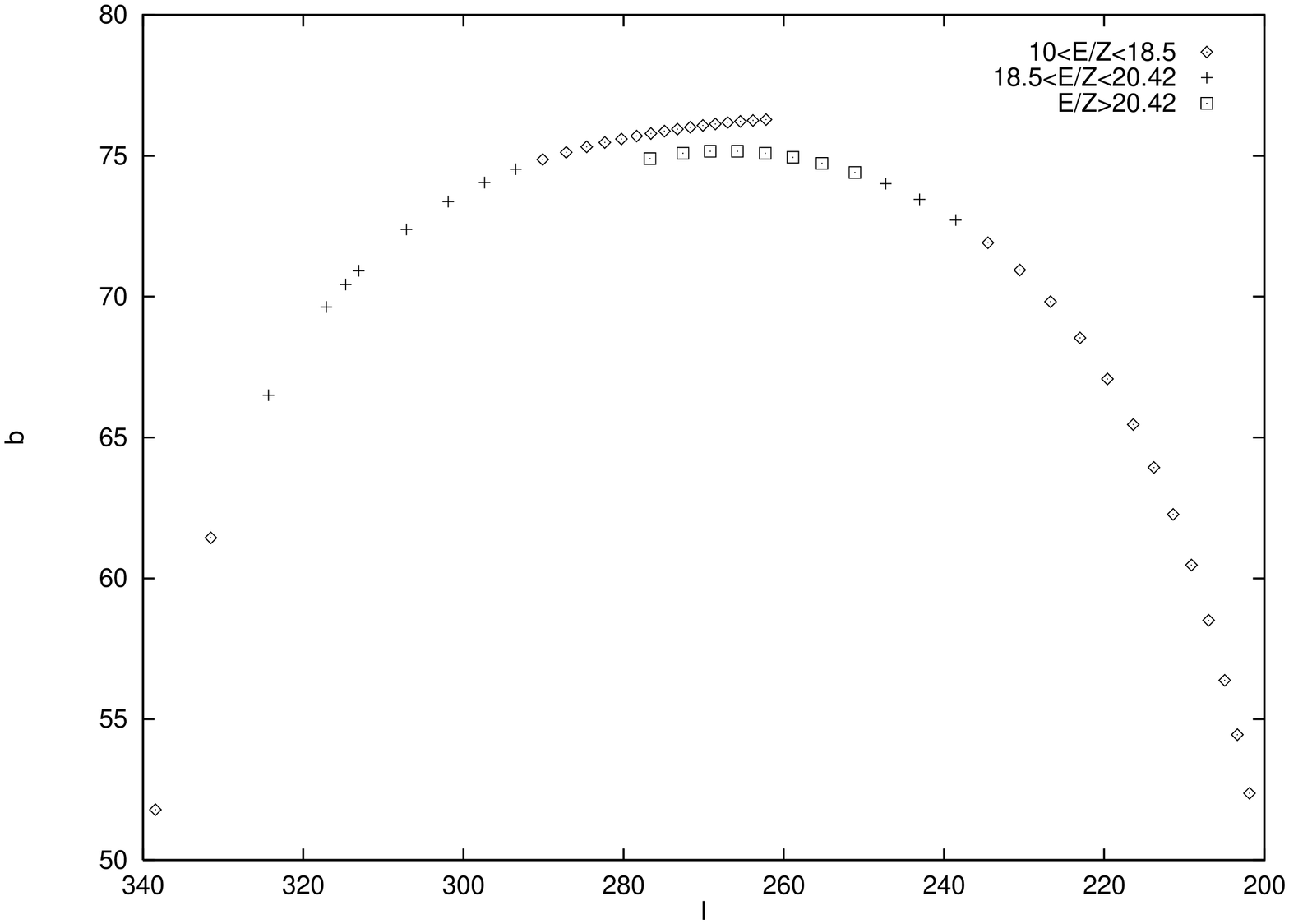,width=6cm,angle=-90}
\epsfig{file=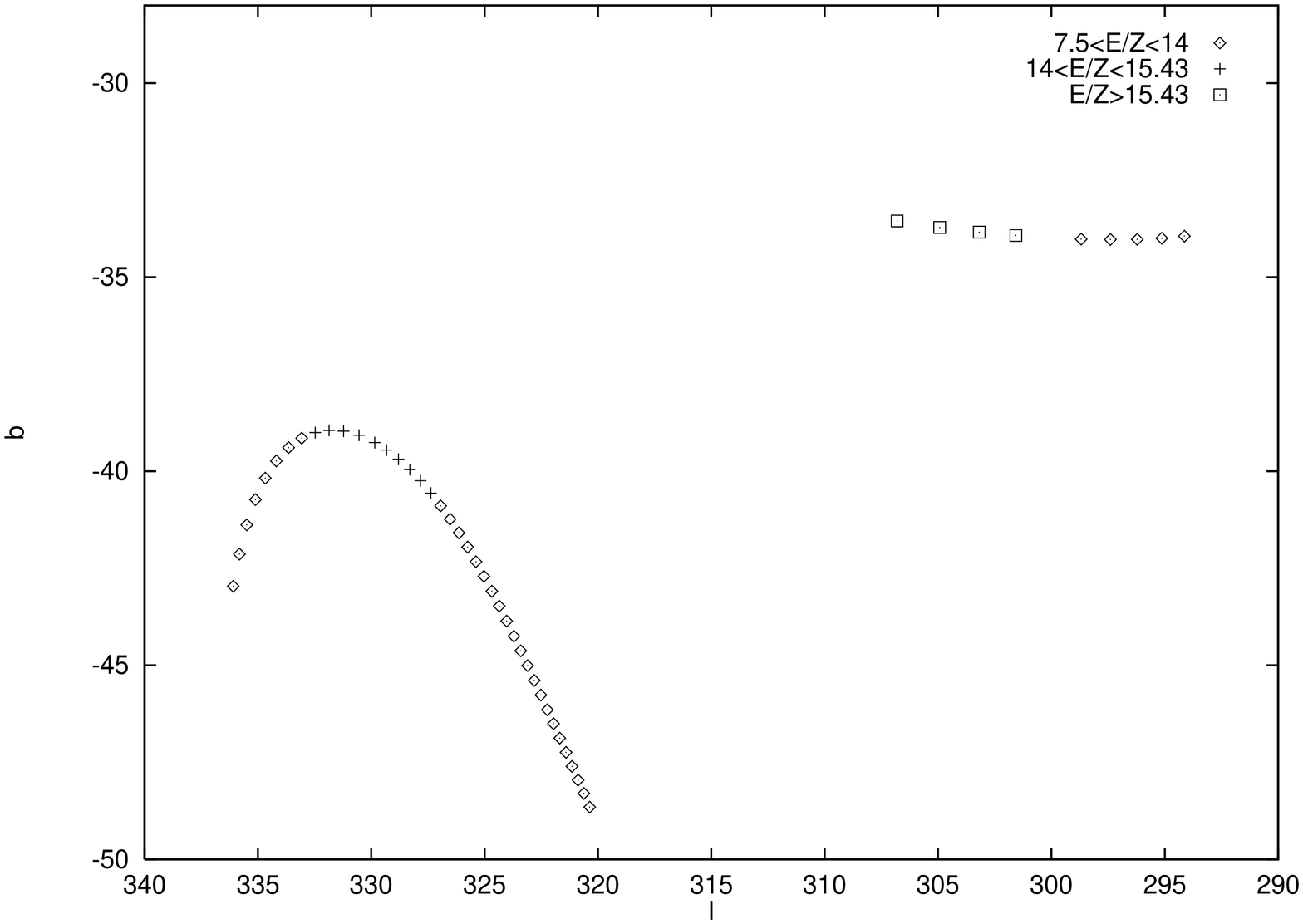,width=6cm,angle=-90}
\epsfig{file=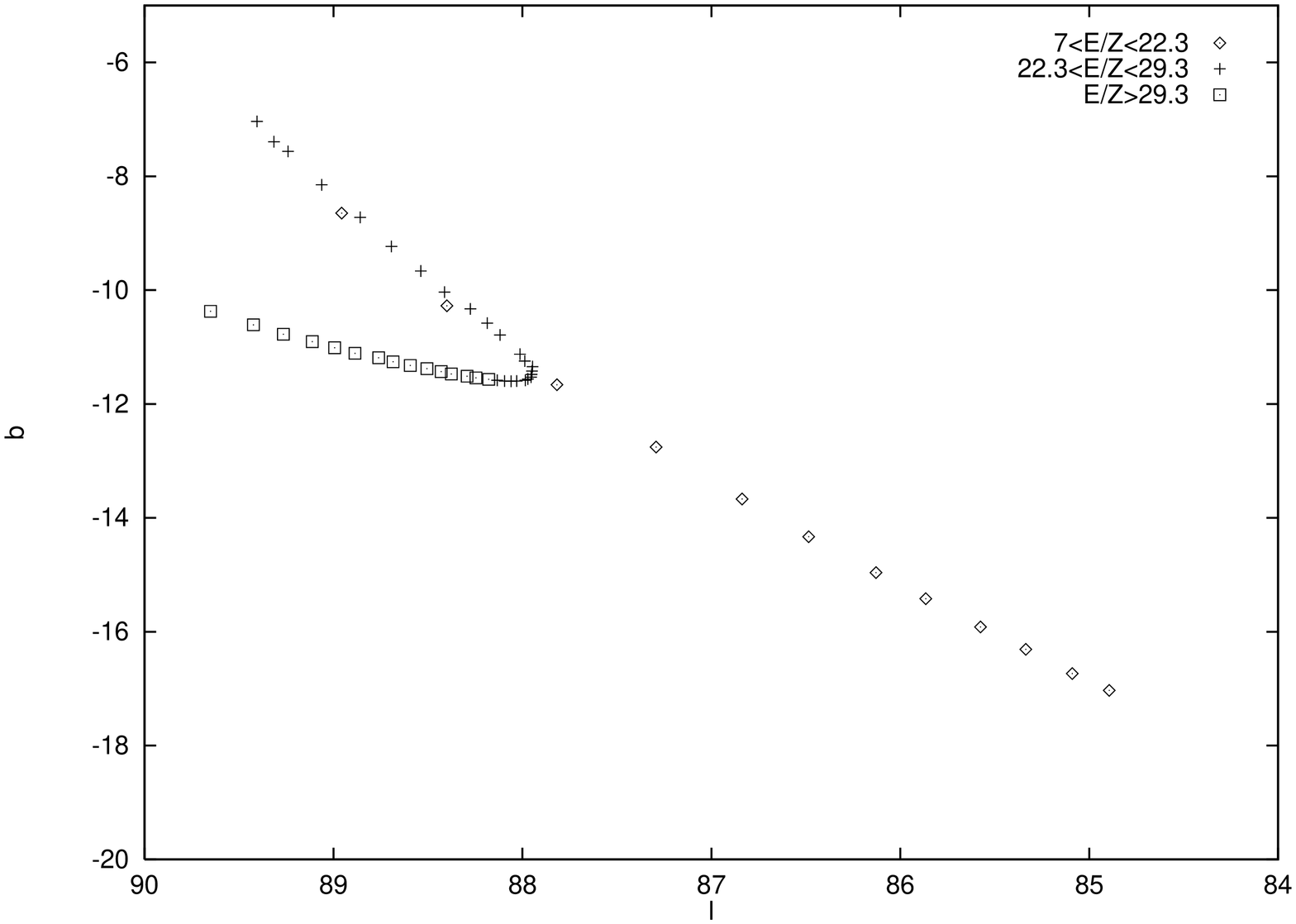,width=6cm,angle=-90}
\caption{Predicted arrival directions of 50 UHECR events from
M87 (top) and from the sources at $(\ell,b)=
(320^\circ,-30^\circ)$ (middle) and  $(\ell,b)=(90^\circ,-10^\circ)$
(bottom), that illustrate the clustering effect due to the existence 
of caustics in the magnetic lens mapping.}}\label{evfig}

In the case of M87 (top panel) 11 events out of a total of 
50  with $E/Z$ larger than 10 EeV  (half the energy of the caustic) 
fall in the narrow energy range between the energy of the caustic 
and just 10\% below ('+' signs).
3 of those 11 events are in the principal image and 8 in the secondary 
images. 
Notice that only 5\% of the events would fall in the same energy 
range if the source were not magnified. 
Notice also that just 8 events correspond to the 
principal image at all energies higher than the energy of the caustic. 
The angular clustering effect for M87 is significant but not
extremely large (in this magnetic field model), partly due to  
the fact that the principal image is also largely magnified around 
the energy of the caustic, and partly because deflections are quite large.

The middle panel in figure~\ref{evfig} displays the effect for the
source located at $(\ell,b)=(320^\circ,-30^\circ)$. In this case 
9 events fall within
10\% of the energy of the caustic in the secondary images, separated by no
more than $5^\circ$. Only 4 events are seen for all energies higher than
the energy of the caustic, scattered over almost $10^\circ$. The events 
at energies below the caustic, down to half its energy, are scattered
over more than $20^\circ$. 

The bottom panel in figure~\ref{evfig} corresponds to the
source located at $(\ell,b)=(90^\circ,-10^\circ)$, which presents
two nearby folds, as depicted in figure~\ref{2foldsfig}. In this case 
nearly half of the events (23) fall within the energy range
between the two caustics $(22.3~{\rm EeV}< E/Z < 29.3~{\rm EeV})$,
while 15 events occur for all higher energies, and just 12 events
appear at lower energies down to $E/Z= 7~{\rm EeV}$. 
The relatively small number
of events in the lower energy range is due to the large deamplification
of the image flux. An unlensed source 
would have instead 86\% of the events in the lower energy range
considered, 9\% in the higher energy range, and just 5\% at the
intermediate energies. Notice that while the arrival directions of the 
events considered are spread over more than $15^\circ$, nearly half
of the events fall within just $2^\circ$ around 
$(l=88^\circ,b=-11^\circ)$.

We stress the fact that six out of the eight 
doublets in the UHECR data listed in table 6 in \cite{uchi99} 
are such that the energies of the events 
in a pair differ by less than 10\%.
The same happens with two events in one of the triplets.
This may be an indication that at least a fraction of
the observed clustering of events may be due to magnetic
lensing around caustics. 

Figure~\ref{evfig} (and figure 6 in paper I) illustrate
the angular displacement of the images as a function of energy.
As discussed in appendix A, the observed angular separation
between images scales as $\sqrt x$, where $x$ measures the
distance from the source to the caustic. Consequently, the
angular separation between the pair of images created at the caustic
crossing scales with
energy near a caustic as 
\begin{equation}
\Delta\theta\approx\Theta\sqrt{1-E/E_0}~.
\label{angle}
\end{equation}
The proportionality constant $\Theta$ is fitted from the
numerical output for the angular displacement of the 
images. In the case of M87, $\Theta\approx 56^\circ$,
and in the case of the source at $(\ell,b)=(320^\circ,-20^\circ)$
it is $\Theta\approx 15.1^\circ$.

\section{Effects upon composition}\label{composition}

Another interesting effect of  magnetic lensing is that it can
strongly alter the measured composition of CRs at high energies.
This is because at each source location the magnification is a
function of $E/Z$. Thus, for a given energy the CR flux of components
with different $Z$ suffers different (de)magnification as the CRs
travel through the Galactic magnetic field.

\FIGURE{
\epsfig{file=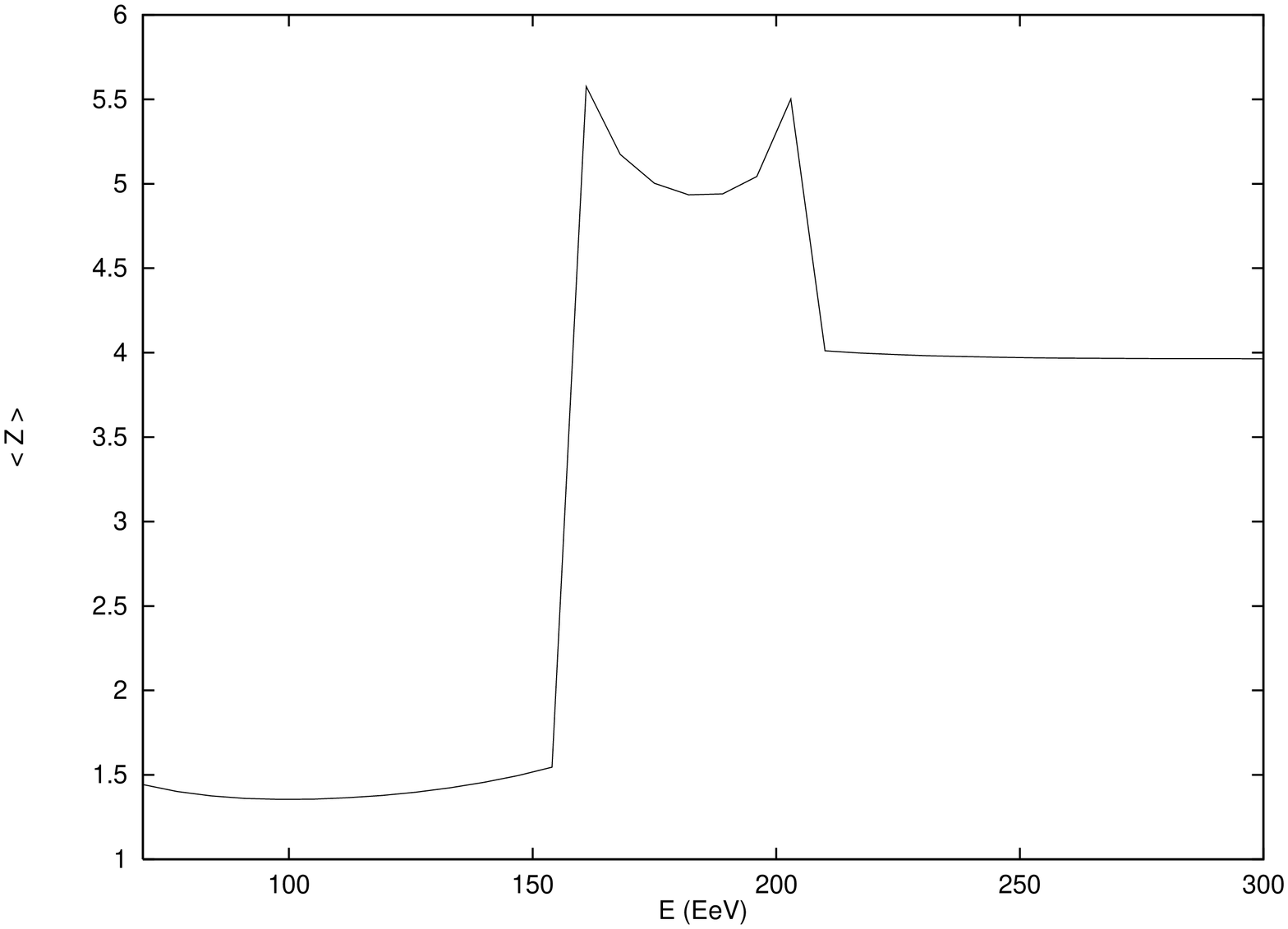,width=8cm,angle=-90}
\caption{Illustration of the effect of magnetic lensing upon the 
UHECR observed composition. The source is located at 
$(\ell,b)=(90^\circ,-10^\circ)$, and  undergoes the magnification 
depicted in figure~\ref{2foldsfig}. The source injects 50\% of 
protons and 50\% of Nitrogen. In the absence of lensing the mean value
of $Z$ would be $\langle Z \rangle = 4$, independently of energy.}}
\label{compofig}

To give an idea of the strength of this effect, let us consider a
simple example: suppose that from a given source the CR flux arriving
to the halo is composed by a fraction $f$ of protons and a fraction
$1-f$ of heavier nuclei with charge $Z_h$. In the absence of
magnetic lensing the mean value of $Z$ of the arriving particles would
be $\langle Z \rangle = f+(1-f)Z_h$. Due to the effect of the magnetic
field, the flux of protons on Earth will be (de)magnified by $\mu (E)$
and the flux of heavier nuclei by $\mu (E/Z)$. Thus, the mean value of
$Z$ of the CRs arriving on Earth will be 
\begin{equation}
\langle Z \rangle =\frac{f \mu(E) +(1-f) Z_h \mu(E/Z_h)}{f \mu(E) 
+(1-f) \mu(E/Z_h)}.
\end{equation}
As the magnifications can be large, $\langle Z \rangle$ can be
strongly modified by the magnetic field. The effect is specially
important for regions of the sky for which the total magnification
$\mu$ (adding the contribution from all the images) has noticeable
changes as $E/Z$ varies.

We show as an example the source located at
$(\ell,b)=(90^\circ,-10^\circ)$, 
which undergoes the magnification depicted in figure~\ref{2foldsfig}. 
We took for reference a flux composed by a fraction $f=0.5$ of protons
and a fraction 0.5 of Nitrogen ($Z_h=7$). 
In the absence of lensing the mean value
of $Z$ would be $\langle Z \rangle = 4 $. We see in figure~\ref{compofig} 
that a clear change from a light composition in the smaller energy region 
to a heavier composition  at the larger energies appears due to magnetic
lensing.

\section{Time delays}\label{td}

A charged UHECR that traverses a distance $L$ within a homogeneous
magnetic field $B$ perpendicular to its trajectory is deflected by an 
angle of the order of  
$\eta\simeq 5^\circ (10~{\rm EeV}~Z/E)(B/\mu{\rm G})(L/ 
{\rm kpc})$, in the limit of small deflections. 
Consider two initially parallel CRs emitted simultaneously, 
one with a much lower ratio $E/Z$ than the other.
If they enter a region permeated by a homogeneous magnetic field and 
converge to the same point after a distance $L$, 
the CR with lower ratio $E/Z$ arrives later, with a relative 
time delay $\delta t\approx 
\eta^2L/2\approx 10~{\rm yrs} 
(10~{\rm EeV}~Z/E)^2(B/\mu{\rm G})^2
(L/~{\rm kpc})^3$. Higher energy events arrive earlier. 

Time delays induced by intergalactic magnetic fields are an essential
ingredient in bursting models for the origin of UHECRs \cite{wax96}.
There are very definite observational signatures of such a scenario.
For instance, individual bursting CR sources would have very narrow
observed spectra, since only CRs with a fixed time delay would be
observed at any given time. The energy at the peak in the 
differential CR flux received on Earth should shift with time as
$t^{-1/2}$. 

Here we analyse the implications of the regular component of the
galactic magnetic field upon time delays between events from a
UHECR source at different energies, and between events from 
different images of a single magnetically lensed source.

Since UHECRs are extremely relativistic, their time delay
compared to straight propagation at the speed of light 
from the source to the
observer is simply determined by the excess path length.
We assume that the extragalactic UHECR sources are sufficiently
distant that we can approximate the CR flux incident upon the
galactic halo as a beam of parallel trajectories, all emitted
simultaneously. We then numerically
determine the difference in path length between the trajectory of a
charged CR and the path length along the parallel straight trajectory 
in the beam that reaches the Earth. In other words, we measure the
delay in the arrival time of a charged CR with respect to the arrival
time of photons (or charged CRs with much higher energy)
emitted simultaneously from a distant source.

Figure~\ref{tdfig}  displays the time delay 
with respect to straight
propagation for the principal and secondary images of the
source locations discussed in the previous sections. 

\FIGURE{\epsfig{file=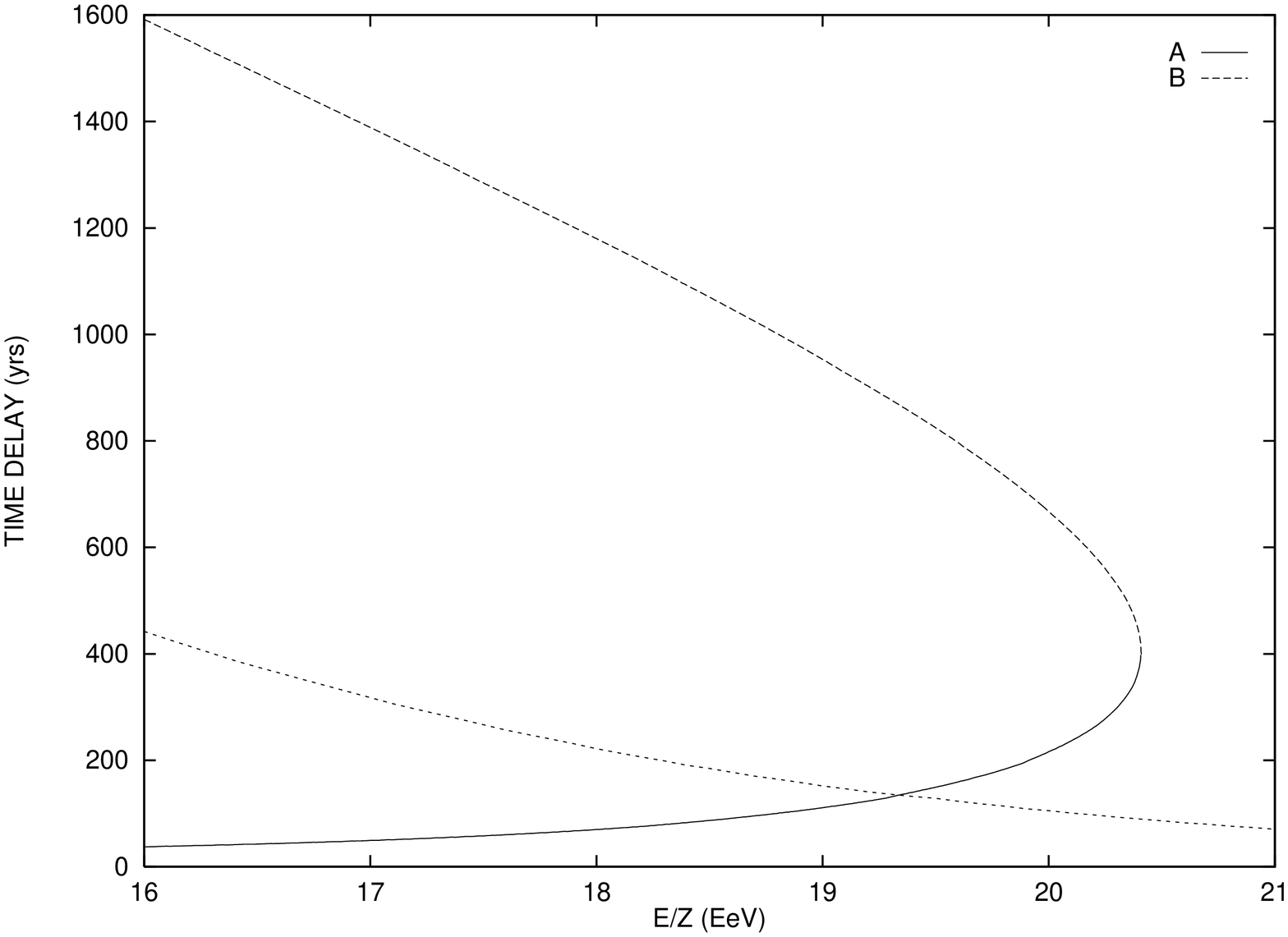,width=6cm,angle=-90}
\epsfig{file=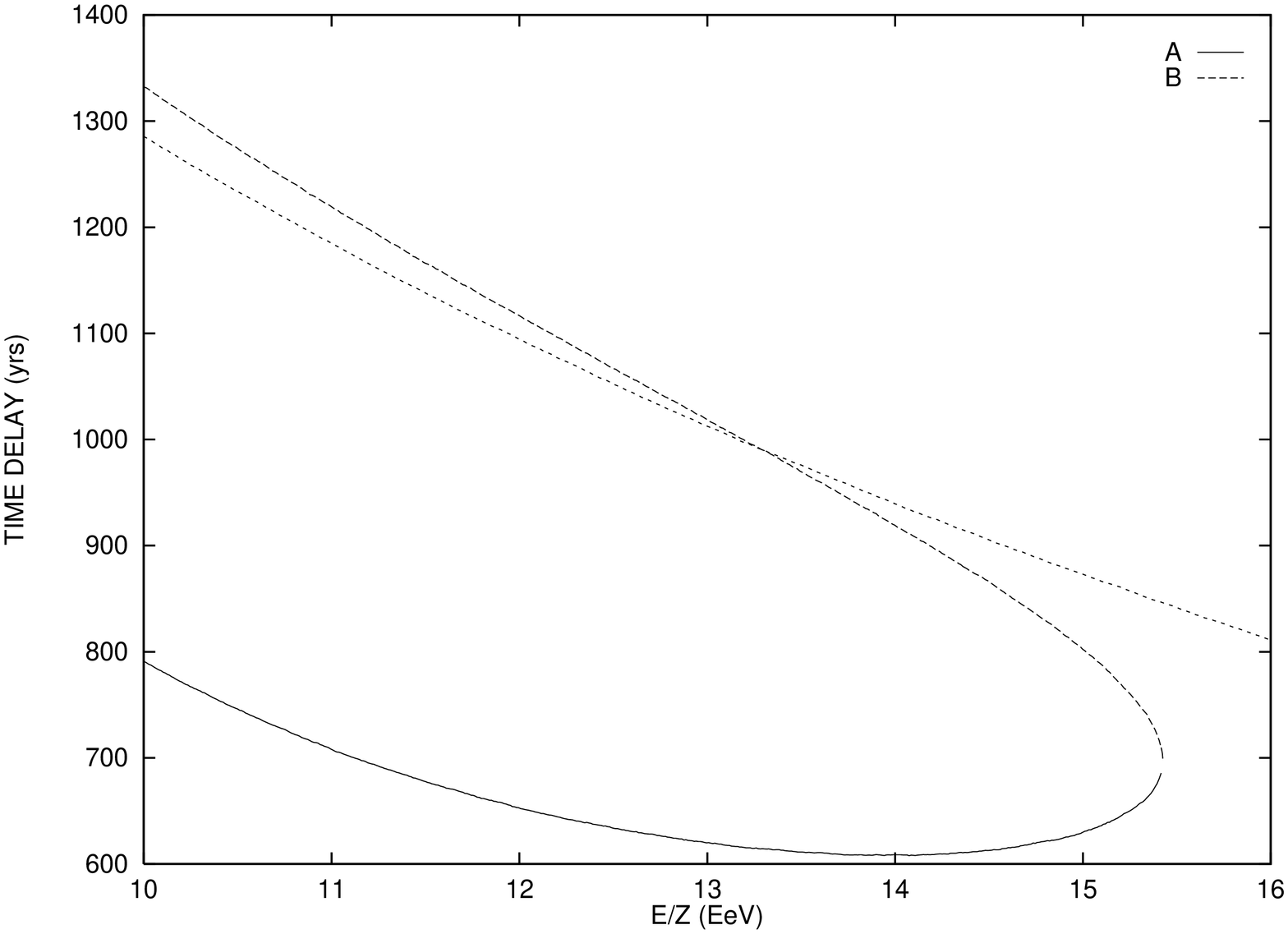,width=6cm,angle=-90}
\epsfig{file=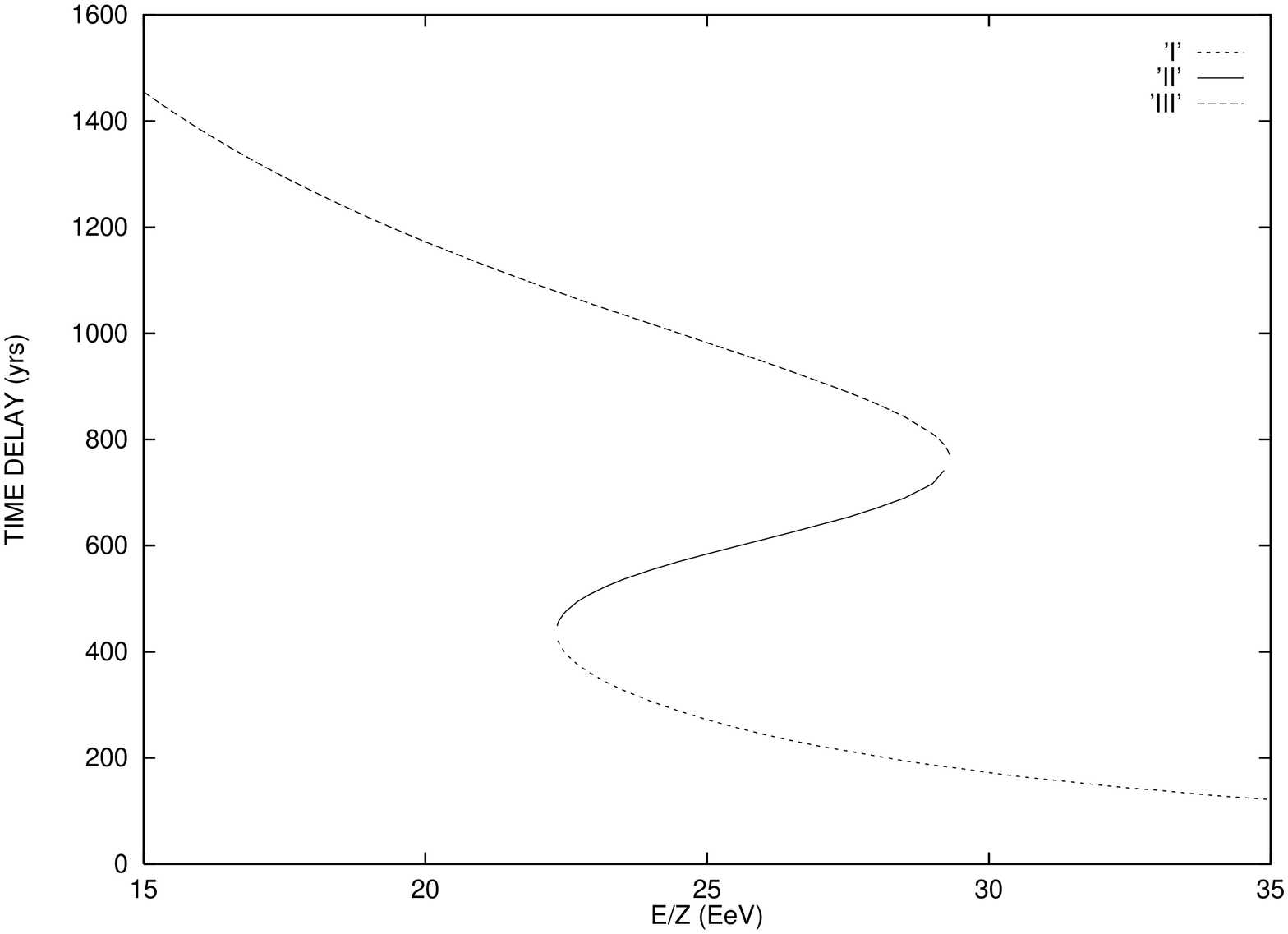,width=6cm,angle=-90}
\caption{Time delays in the arrival of UHECRs from 
the principal and secondary images a distant
source, measured with respect to the arrival time of a photon
(or a much more energetic charged CR). The relative arrival time
delay of a pair of images near the caustic scales as
$\delta t\propto \sqrt{1-E/E_0}$. Notice that the arrival time
of one of the images {\bf  decreases} for decreasing $E$
below the energy caustic, opposite to the behaviour in normal 
regions.} 
}\label{tdfig}

The time delay in normal regions (far from critical points) 
monotonically increases with decreasing energy. The spiraled
galactic magnetic field is far from homogeneous, and thus 
departures from the $E^{-2}$ dependence are certainly expected.
We have checked that the time delay averaged over a regular 
grid of arrival directions scales, in the BSS-S galactic 
magnetic field
model considered, as $<\delta t>\approx 1000~{\rm yrs} 
(10~{\rm EeV}~Z/E)^2$ from very high energies down to 
$E/Z$ of order  5 EeV.  At lower $E/Z$ values the
increase in the time delay with decreasing energy is faster, 
since a sharp transition from quasirectilinear to drift motion
occurs at values of $E/Z$  between 3 and 1 EeV.

What strikes the eye in figure~\ref{tdfig} 
is that the time delay of one of the images in a pair can have an
energy dependence opposite to that in normal regions.
Indeed, the time delay of one of the images that are visible
at energies below the energy of the caustic increases with decreasing
energy, while the time delay of the other member of the pair decreases. 
Thus, the relative arrival time of events from a single 
image of a CR source does not necessarily increase with decreasing 
energy. It is often argued that the doublets in which the 
highest energy event arrived later than the other member in the pair 
can not arise from bursting sources. As we have seen, 
this is not necessarily true 
near a caustic.

The relative arrival time delay between equal energy events from 
different images $A$ and $B$
in a pair behaves near the caustic as
\begin{equation}
\Delta t=\delta t_A
-\delta t_B\approx T\sqrt{1-E/E_0}~.\label{deltat}
\end{equation}
This can be understood as follows, in the limit of small deflections.
The trajectory of each of the images in the pair is deflected by
$\eta$ and $\eta +\Delta\theta$ respectively, with $\Delta\theta$
given by eq.~(\ref{angle}). The time delays
with respect to straight propagation are thus proportional to
$\eta^2$ and $\eta^2 +2\eta\Delta\theta$ respectively,
as long as $\Delta\theta\ll\eta$. 
Thus the relative
time delay between events at the same energy from the two images
scales as $\Delta\theta$ at energies sufficiently close to that of
the caustic.
The fit of eq.~(\ref{deltat}) to the numerical output is highly accurate, 
with $T\approx  
3200$~yrs in the case of M87  
and $T\approx  1030$~yrs in the case of the source in the southern 
sky at $(\ell=320^\circ,b=-30^\circ)$.

\section{Conclusions}\label{conclusions}

UHECRs beyond the ankle in the spectrum are most probably nuclei of
extragalactic origin. In their way
to the Earth they feel the magnetic field structure permeating the
Galaxy and hence their trajectories become deflected and their fluxes are
lensed.  The implications of this magnetic lensing are manyfold and
they have to be taken into account in the analysis of the
observations. 

Since the lensing effect depends on the energy it can sizeably affect
the observed spectrum of the sources. Moreover, for sources located
in a large fraction of the sky CRs can reach the Earth following
different paths, and hence multiple images of those sources will
appear. A useful way to visualise these effects is to display the
mapping of the arrival
directions in the Earth into the incoming directions outside the Galaxy,
as was introduced in paper I with what was dubbed a `sky sheet'.
In the locations where this surface develops folds, pairs of
additional images of the source are present. These folds,
corresponding to the  caustics, move with energy, and as they cross a
given source location pairs of additional images appear or
disappear. Near the fold the magnification of each image in the pair
diverges as $\mu\propto 1/\sqrt{|E-E_0|}$. Thus the probability to
detect events from a given source is noticeably enhanced for energies
close to $E_0$,  at which the caustic crosses the source location.
This also leads to an expected concentration of events near the
location at which the new pair of images appears. This is relevant in
the analysis of the small scale clustering present in the UHECR 
distribution and in this respect it is remarkable that the observed
events in doublets and triplets in most cases are very close in
energies. With the increased statistics expected with the new
detectors, such as Auger \cite{auger} or High Res \cite{hires}, 
these features may become
testable through a careful analysis of the clustering of the events. 

Also the observed CR composition can be affected by magnetic lensing 
due to the dependence of the flux amplification on $E/Z$. Nuclei with 
different charges are magnified by different amounts for a given energy.  
This effect is sizeable for sources whose magnification has a strong
energy dependence (in particular when there are caustics) and which
have a mixed composition.

Another feature which we have discussed is the time delay due to the
galactic field between the different images of a lensed source. These
are typically larger than the lifetimes of the CR observatories, and are
strongly dependent on the energy.
If burst sources exist, narrow spectra will then result at any given
time, and the
different images will be observed simultaneously with different $E/Z$
values. 

We have to stress that in addition to the effects related to the
magnetic field of the Galaxy discussed in this paper there may be
also similar effects associated to the magnetic field in the source
galaxy or even to the intergalactic fields if these ones are
strong. Also the magnetic field model adopted here is plausible but the
real one may differ from it, changing the quantitative  details but
not the general qualitative results.

\acknowledgments

Work partially supported by ANPCyT, CONICET and Fundaci\'on Antorchas, 
Argentina.

\appendix

\section{Magnification near caustics}

The relation between the source position and the image positions is
given by a mapping from the source coordinates $(\ell,b)
_{H}$ to the observer's ones $(\ell,b)
_E$. When this mapping is multiple valued
there will be additional image pairs appearing (see paper I). The
magnification of a given image will be $\mu={\rm d}\Omega_H/$d$\Omega_E$,
i.e. the ratio of differential solid angles associated to the
mapping. 

To visualise these facts it is useful to consider the inverse map $(\ell,b)
_E\to (\ell,b)_{H}$  and look at it as the mapping of a surface (the
observer's sky) into another surface (dubbed the `sky sheet' in I),
which will be folded when multiple images appear, and whose stretching
is related to the magnification of the images. The location of the
folds correspond to the caustics along which image pairs with
divergent magnification appear and the magnification varies very
rapidly with the distance to the fold. Since the folds move as the
energy is decreased, the knowledge of the magnification as a function
of the angular distance to the fold for a given energy can be used to
obtain the magnification, for a given source,  as a function of the
energy near a caustic, which is the quantity of interest to us here.

The angular dependence of the magnification near a fold has a very
simple geometrical interpretation in terms of the folded sky
sheet. Let us take local angular
coordinates $(x,y)$ in the source sky such that $x=0$ describes
locally the location of the fold (and hence the fold is along the $y$
axis while the $x$ axis is orthogonal to it). We can also adopt
(non--orthogonal) local coordinates on Earth $(X,Y)$ such that $X=0$
is mapped into $x=0$ and similarly $Y=0$ is mapped into
$y=0$. Furthermore, we will assign a third coordinate to the mapping,
i.e. $(X,Y)\to (x,y,z)$, giving a depth to the sky sheet so that the
fold can be visualised. The simplest choice for this (arbitrary) third
coordinate is to take $z=X$, as we will do in the following. The
results will be easier to obtain with this choice, but are independent
of it. In this way the mapping for a fixed value of $Y$ in a
neighbourhood of $X=0$ will look as
shown in figure 6. The magnification is given by
$\mu=|K {\rm d}X/{\rm d}x|$, where $K\equiv [\cos
b_H|\partial(\ell,b)_H/ \partial(x,y)|]/[\cos
b_E|\partial(\ell,b)_E/ \partial(X,Y)|] {\rm d}y/{\rm d}Y\simeq
K_0(Y)+K_1(Y)X+O(X^2)$.
The next step is to relate the factor d$X$/d$x={\rm d}z/$d$x$
with  the slope  of the
curve in figure 6, 
and approximate the fold near the caustic by its
Taylor expansion $x=a z^2+bz^3+O(z^4)$. Hence we have 
\begin{equation}
{{\rm d}x\over {\rm d}z}\simeq 2az+3bz^2
\end{equation}
and hence, since $z=\pm \sqrt{x/a}(1+O(x))$, with the sign indicating
to which side of the fold the image belongs, we have 
\begin{equation}
\mu={A'\over\sqrt{x}}\pm B'+O(\sqrt{x}),
\end{equation}
$A'$ and $B'$ being $x$-independent parameters 
related to the parameters $a$, $b$, $K_0$ and $K_1$.

\FIGURE{\epsfig{file=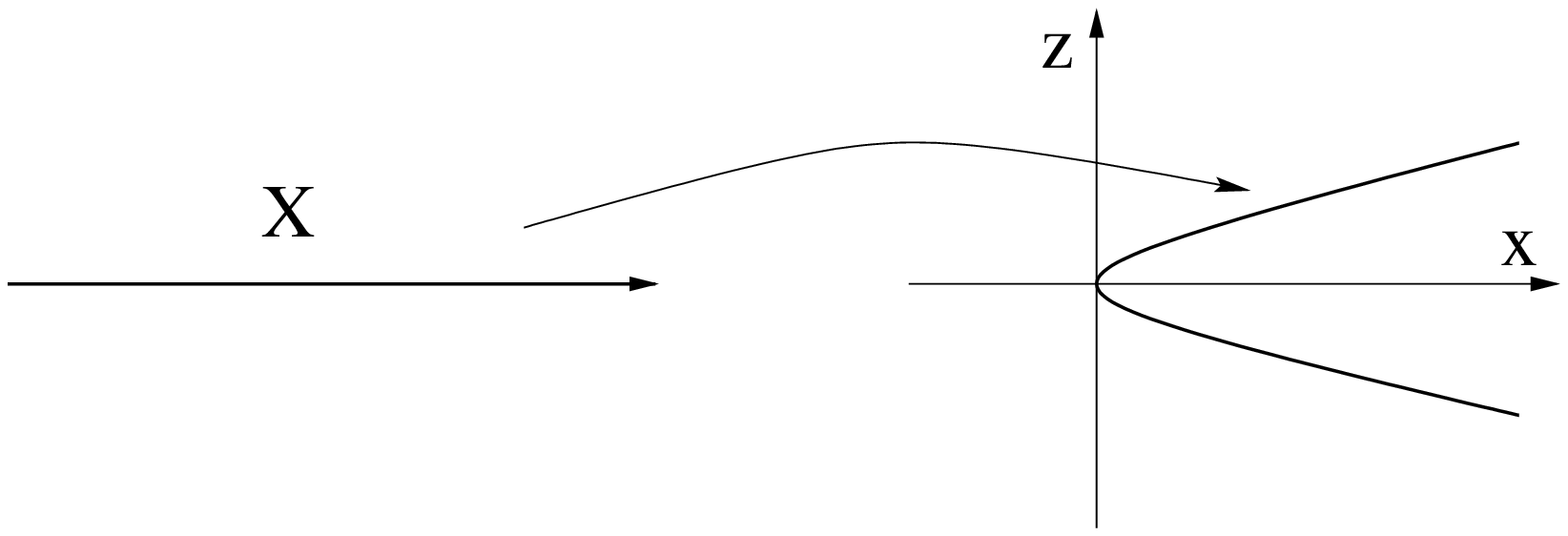,width=4cm,angle=-90}
\caption{Mapping from the observer's sky ($X$) into the source sky
($x$), for fixed $Y$, 
with the vertical coordinate ($z=X$) allowing to visualise the
fold in the sky-sheet.}}\label{figa1}

The final step is to relate this with the magnification for one given
source as a function of energy. Suppose that at energy $E_0$ a source 
located at $(\ell_0,b_0)
_{H}$ would be just on top of the fold. 
If we define the angular `velocity' of the fold 
as the energy is changed as $V(E)\equiv {\rm d}x_f/{\rm d}E$, with
$x_f$ being  the distance 
from the source to the fold ($x_f=0$ for $E=E_0$),
one has $x_f\simeq V(E_0)|E_0-E|$. Hence we finally get
\begin{equation}
\mu(E)={A\over \sqrt{1-E/E_0}}\pm B+O(\sqrt{1-E/E_0}).
 \end{equation}

\section{Magnification for two nearby folds}

The previous Appendix dealt with the magnification for the two images
which appear when a fold crosses the source location. The third image
(e.g. the one present originally) was assumed to be far from the fold
so that its magnification was supposed to have a non-singular
behaviour. 
Another situation of interest is when two folds are nearby (e.g. when
the sky--sheet has a narrow fold which moves with energy across the
source, rather than becoming wider and remaining on top of the source,
or when the source is near a cusp where two folds merge) so that the
three images can have large magnifications.

When a couple of nearby folds crosses a source, we see that 
 a pair of
images appear at some energy and at a somewhat smaller energy one of
the new images merges with the original one present at high energies
and they disappear (figure~\ref{2foldsfig}).
In this case there is a relation between the magnification of the
three images, as we now show. A similar relation holds in the
gravitational lens case \cite{sch92b}. 

Following similar lines as in the
previous Appendix, the folded surface can be described as in
figure 7. 
Without loss of generality we can choose the origin $x=0$ so that the
vertical coordinates of the other two images at $z_1$ and $z_2$ in
figure~\ref{figa2} 
satisfy $z_2=-z_1$. In this case a Taylor expansion of the
fold will be $x=az^3-cz$ (with $a,c>0$ and no quadratic term due to
our choice of origin for $x$). The location of the folds are at
$x_I=-2a(c/3a)^{3/2}$ and $x_{II}=-x_I$. The images in region I
($z>z_I=\sqrt{c/3a}$) and III ($z<z_{II}=-z_I$) will have positive
parities, while the one in region II ($z_{II}<z<z_I$) will have
negative parity.

\FIGURE{\epsfig{file=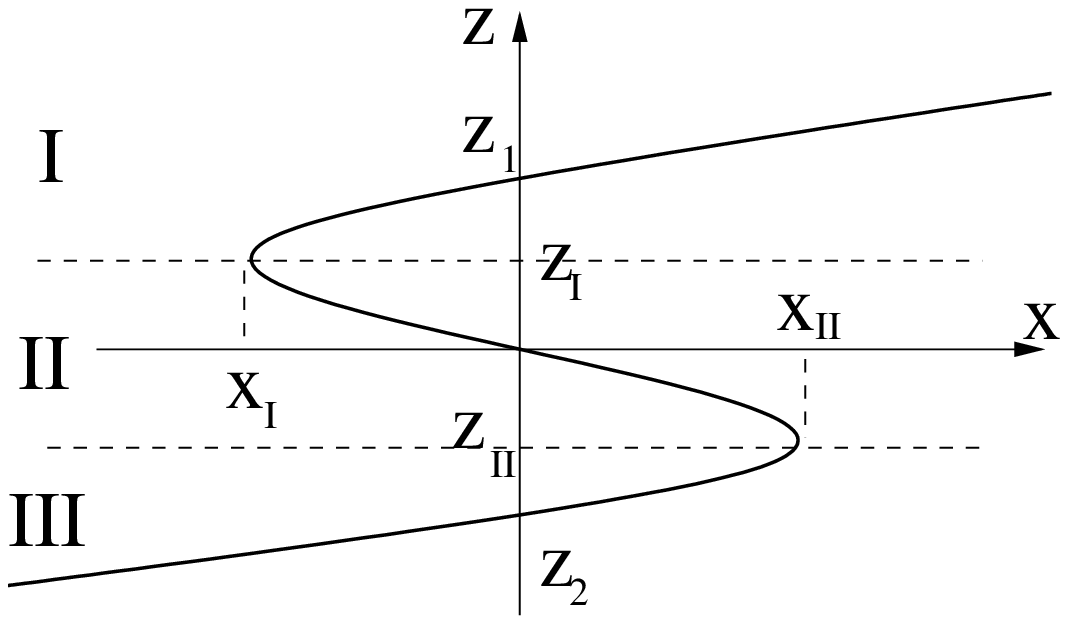,width=6cm,angle=-90}
\caption{Sky-sheet in the source sky corresponding to two nearby
folds (for $Y$ fixed).}}\label{figa2}

The vertical coordinates of the three images for a given $x$ are
simply obtained as
\begin{equation}
z_k=2z_I\cos{\alpha+2k\pi\over 3}\ \ \ \ \ (k=0,1,2)
\end{equation}
where $\cos\alpha=x/x_{II}$.
Here $k=2$ corresponds to the negative parity image (region II) while
$k=0$ and 1 are the images in region I and III respectively.

The magnification of the images will be $\mu_k=K|{\rm d}z_k/{\rm d}x|$,
where 
\begin{equation}
{{\rm d}z_k\over {\rm d}x}={2z_I\over 3x_{II}\sqrt{1-(x/x_{II})^2}}
 \sin\left[{\cos^{-1}(x/x_{II})\over 3}+{2k\pi\over
3}\right].
\end{equation}
From this we find that
\begin{equation}
\left| {{\rm d}z_0\over {\rm d}x}\right| +
\left| {{\rm d}z_1\over {\rm d}x}\right| =
\left| {{\rm d}z_2\over {\rm d}x}\right|\, ,
\end{equation}
i.e. that the sum of the magnifications of the two positive parity
images coincides with the magnification of the negative parity image.

Next to leading terms (i.e. corrections of $O(\sqrt{x})$ will add
constant terms to the magnifications, so that the final result up to
corrections of $O(x)$ is
\begin{equation}
\mu_I+\mu_{III}=\mu_{II}+{\rm const}.
\end{equation}
This theorem is illustrated in figure~\ref{2foldsfig}.

\end{document}